\documentclass[aps,prd,showpacs,amsmath,10pt,twocolumn,superscriptaddress,floatfix,nofootinbib,notitlepage]{revtex4-2}
\usepackage[utf8]{inputenc}
\usepackage[english]{babel}
\usepackage[dvipsnames]{xcolor}
\usepackage[normalem]{ulem}
\usepackage[
colorlinks=true,
linkcolor=blue,
breaklinks=true,
urlcolor=magenta,
citecolor=blue]{hyperref}
\usepackage{footmisc}
\usepackage{bm,color,xcolor,amsfonts,amsmath,amssymb,epsfig,esint,orcidlink}

\usepackage[utf8]{inputenc}
\usepackage{slashed}
\usepackage{amsmath}
\usepackage{amssymb}
\usepackage{gensymb}
\usepackage{color}
\usepackage{hyperref}
\usepackage{xcolor}
\usepackage{slashed}
\usepackage{braket}
\usepackage{bm}
\usepackage{cancel}

\usepackage{indentfirst}
\allowdisplaybreaks

\newcommand{\dszero}{D^*_{s0}}
\newcommand{\dsone}{D_{s1}}

\newcommand{\be}{\begin{equation}}
\newcommand{\bea}{\begin{eqnarray}}
\newcommand{\ee}{\end{equation}}
\newcommand{\eea}{\end{eqnarray}}

\newcommand{\Br}{\mbox{Br}}
\newcommand{\kappaon}{\kappa_{\rm loop}(q^2=m_{\dszero}^2)}
\newcommand{\kappaoff}{\kappa_{\rm loop}(q^2)}

\graphicspath{{Figs.dir/}}

\newcommand{\fzj}{\affiliation{Institute for Advanced Simulation, Forschungszentrum J\"ulich, D-52425 J\"ulich, Germany}}

\newcommand{\itp}{\affiliation{Institute of Theoretical Physics, Chinese Academy of Sciences, Beijing 100190, China}}

\newcommand{\ucas}{\affiliation{School of Physical Sciences, University of Chinese Academy of Sciences, Beijing 100049, China}}

\newcommand{\hiskp}{\affiliation{Helmholtz-Institut f\"ur Strahlen- und Kernphysik% %and Bethe Center for Theoretical Physics
, Universit\"at Bonn, 53115 Bonn, Germany}}

\begin{document}

\title{What can we learn from the radiative decays of the $\dsone(2460)$ meson?}

\author{Hai-Long Fu\orcidlink{0000-0002-1722-4145}}\email{fuhailong@itp.ac.cn}
\itp \ucas

\author{Feng-Kun~Guo\orcidlink{0000-0002-2919-2064}}\email{fkguo@itp.ac.cn}
\itp \ucas %\peng \scnt

\author{Christoph~Hanhart\orcidlink{0000-0002-3509-2473}}\email{c.hanhart@fz-juelich.de}
\fzj

\author{Alexey Nefediev\orcidlink{0000-0002-9988-9430}}\email{a.nefediev@uni-bonn.de}
\hiskp
%\lis

\begin{abstract}
We study the radiative decays $\dsone(2460)\to\gamma D^{*}_{s0}(2317)$ and $\dsone(2460)\to \gamma D^0K^+/\gamma D^+K^0$ and argue that their simultaneous experimental measurement, or at least a constraint on the ratio of the corresponding branching fractions, can allow one to probe the nature of the $D^{*}_{s0}(2317)$ and $\dsone(2460)$ mesons.
\end{abstract}

\maketitle

\section{Introduction}

Radiative decays play a distinguished role in strong interaction physics as they often provide a convenient doorway to establishing the nature of hadronic states. The photon emission vertex is controlled by Quantum Electrodynamics (QED)---the most developed and well-understood field theory within the Standard Model.
The small QED coupling constant, $\alpha_{\rm em}=e^2/(4\pi)\approx 1/137$, results in the suppression of electromagnetic radiative processes by two orders of magnitude in probability compared to analogous nonradiative reactions.
However, if the fine structure curse is overcome by the experimental statistics, then the information gained on the studied hadronic states may provide a valuable reward for the efforts spent.

It should be noted that the radiative decays of different hadronic systems are sensitive to different components of their wave functions and, therefore, provide different insights into the nature of such states. For example, the radiative decays $\phi(1020)\to\gamma S$ and $S\to\gamma V$, with $S$ for $a_0/f_0(980)$ and $V$ for $\rho$, $\omega$, $\gamma$, exhibit quite distinct hierarchy patterns for compact or molecular structures of the scalar mesons~\cite{Kalashnikova:2005zz}. 
In contrast, the experimentally measurable ratio of the radiative decay widths $\Gamma(X(3872)\to\gamma\psi(3686))/\Gamma(X(3872)\to\gamma J/\psi)$~\cite{BaBar:2008flx,Belle:2011wdj,LHCb:2014jvf,BESIII:2020nbj,LHCb:2024tpv} is sensitive to the short-range component of the $X$ wave function and appears not to be decisive concerning its molecular component~\cite{Guo:2014taa}; see also Ref.~\cite{Guo:2026ngz} for a pedagogical 
introduction to the subject.
The $P$-wave positive-parity $D_{sJ}$ mesons, with $J=0,1,2$, offer yet another example of hadronic systems whose radiative decays may serve as a probing tool for investigating their nature.
A recent update on the widths of the radiative decays $\dszero(2317)\to\gamma D_s^*$ and $\dsone(2460)\to\gamma D_s^{(*)}$ evaluated in the molecular model for the decaying $D_{sJ}$ mesons can be found in Ref.~\cite{Fu:2021wde} (see also Refs.~\cite{Lutz:2007sk,Cleven:2014oka} for earlier studies in the same spirit).
These decays have also been comprehensively calculated in the chiral doublet model, which assumes that the $\dszero(2317)$ and $\dsone(2460)$ mesons are chiral partners of the pseudoscalar $D_s$ and vector $D_s^*$ mesons, respectively, with opposite parity~\cite{Bardeen:2003kt}.
A calculation of the radiative decay widths $\dsone(2460)\to\gamma D_s^{(*)}$ and $\dsone(2536)\to\gamma D_s^{(*)}$ performed assuming all involved charm-strange mesons to be conventional quark--antiquark states is presented in Ref.~\cite{Bondar:2025qzg}. In Ref.~\cite{Zhang:2024usz}, the decays $\dsone(2460)\to\gamma D_s^{(*)}$ are studied
in a model that includes both $c\bar s$ and
two-hadron components. Since the existing experimental information on such radiative decays is very limited, further experimental studies as well as reanalyses of the already existing data from different collaborations in the spirit of Ref.~\cite{Bondar:2025gsh} are necessary to shed light on the nature of these hadronic states. The results on $\dszero\to \gamma D^{*}_{s}$ and $D_{s1}\to\gamma D_{s}$ radiative decays from the perspective of the QCD sum rule can be found in Ref.~\cite{Wang:2006mf}.

\begin{figure*}[t]
\centering
\includegraphics[width=0.8\textwidth]{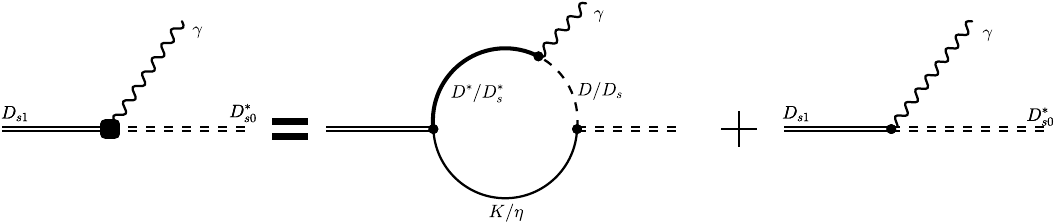}
\caption{The loop and contact contributions to the decay amplitude $\dsone(2460)\to \gamma \dszero(2317)$.}
\label{fig:Ds1Ds0gamma}
\end{figure*}

In this work, we focus on yet another radiative decay, $\dsone(2460)\to \gamma \dszero(2317)$, which can be employed to probe the nature of the involved $D_{sJ}$ mesons.
There are two kinds of contributions possible to this decay that are depicted in Fig.~\ref{fig:Ds1Ds0gamma}. 
The loop diagram is sensitive to the molecular component of the $D_{sJ}$ mesons while
the second diagram (with a contact term)
depends on an a priori unknown parameter, hereinafter denoted as $\kappa_{\rm cont}$, that describes the contribution from short-range physics~\cite{Lutz:2007sk,Cleven:2014oka,Fu:2021wde}. We argue in this work that since a sufficiently accurate experimental measurement of the partial decay width $\Gamma(\dsone\to \gamma \dszero)$ is currently not available \cite{ParticleDataGroup:2024cfk}, an alternative experimental input may be useful to quantify the aforementioned short-range term
and in this way get access to the significance of the loop contribution. In particular, we also study the three-body radiative decays $\dsone(2460)\to \gamma D^0K^+$ and $\dsone(2460)\to \gamma D^+K^0$ and argue that they can be useful in this context. In particular, we demonstrate that the ratio of branching fractions,
\be
{\mathcal R}=\frac{\Br(\dsone(2460)\to \gamma D^{*}_{s0}(2317))}{\Br(\dsone(2460)\to \gamma D^0K^+)},
\label{ratioR}
\ee
is very sensitive to the value of the short-range parameter $\kappa_{\rm cont}$ and thus can be used to determine or at least strongly constrain it. In this way, it should become possible to
improve our understanding of the nature of the $\dszero(2317)$ and $\dsone(2460)$ mesons and make precise theoretical predictions for various reactions involving the vertex $\dsone\dszero\gamma$ as a building block or related to it via heavy-quark symmetry.

 We notice that the studied radiative decays of the $\dsone(2460)$ naturally fill a gap between the two limiting cases previously addressed in the literature. Indeed, on the one hand, the short-range contact term is absent in the radiative decays $\phi(1020)\to\gamma S$, with the scalar mesons $S=a_0/f_0(980)$ in the final state treated as $K\bar{K}$ molecules, so the decay amplitudes can be expressed solely through the $S\to K\bar{K}$ vertex function \cite{Close:1992ay,Achasov:1996ei,Kalashnikova:2004ta}. On the contrary, the radiative decays $X(3872)\to\gamma\psi$, with $\psi=J/\psi$, $\psi(2S)$, are dominated by the short-range component of the $X$ wave function \cite{Guo:2014taa}. Furthermore, one as a matter of principle cannot employ the data on these radiative decays to quantify this component in a model-independent way.  As will be argued below, the radiative decays of the $\dsone(2460)$ addressed in this work lie between the two aforementioned cases since they are sensitive to the short-range contribution to the $\dsone$ wave function and the latter can in principle be quantified using the experimental data. This theoretical insight and the 
estimates made for the already existing and anticipated  experimental measurements constitute the main result of this work.

The paper is organised as follows: In Sec.~\ref{sec:2body}, we evaluate the width of the two-body radiative decay $\dsone(2460)\to \gamma \dszero(2317)$ and study its dependence on the short-range parameter $\kappa_{\rm cont}$. We provide estimates for the value of this parameter based on existing theoretical predictions and experimental data. In Sec.~\ref{sec:3body}, we calculate the
widths of the three-body radiative decays $\dsone(2460)\to \gamma D^0K^+/\gamma D^+K^0$, which also depend on $\kappa_{\rm cont}$ through the intermediate vertex $\dsone\dszero\gamma$. In Sec.~\ref{sec:discussion}, we discuss the dependence of the ratio in Eq.~\eqref{ratioR} on $\kappa_{\rm cont}$ and argue that conclusions about the nature of the $\dszero(2317)$ and $\dsone(2460)$ mesons can be drawn from experimental measurements of this ratio.

\section{The decay \texorpdfstring{$\dsone(2460)\to \gamma \dszero(2317)$}{Ds1(2460)toDs0gamma}}
\label{sec:2body}

The amplitude of the radiative decay $\dsone(2460)\to \gamma\dszero(2317)$ can be expressed as
\be
\mathcal{M}(\dsone\to\gamma\dszero)
=2\kappa(q^2)\varepsilon_{\mu\nu\alpha\beta}\;p_3^\mu \epsilon^{*\nu}(p_3)\epsilon^\alpha(P)v^\beta,
\label{M1}
\ee
with $\kappa(q^2)$ for the transition amplitude (here $q=P-p_3$), $P$ and $p_3$ for the 4-momenta of the $\dsone$ meson and photon, respectively, and the corresponding $\epsilon$'s for their polarisation vectors. The decay width is then calculated as
\be
\Gamma(\dsone\to\gamma \dszero)=\frac{\kappa(m^2_{D_{s0^*}})^2\omega^3}{3\pi m_{\dsone}^2},
\label{WidthDs1Gsogamma}
\ee
where $\omega\approx 139$~MeV is the energy of the photon in the rest frame of the decaying particle and $\kappa(q^2)$ is evaluated at $q^2=m_{\dszero}^2$ for the on-shell $\dszero$ meson.

\begin{figure}
\centering
\includegraphics[width=0.99\linewidth]{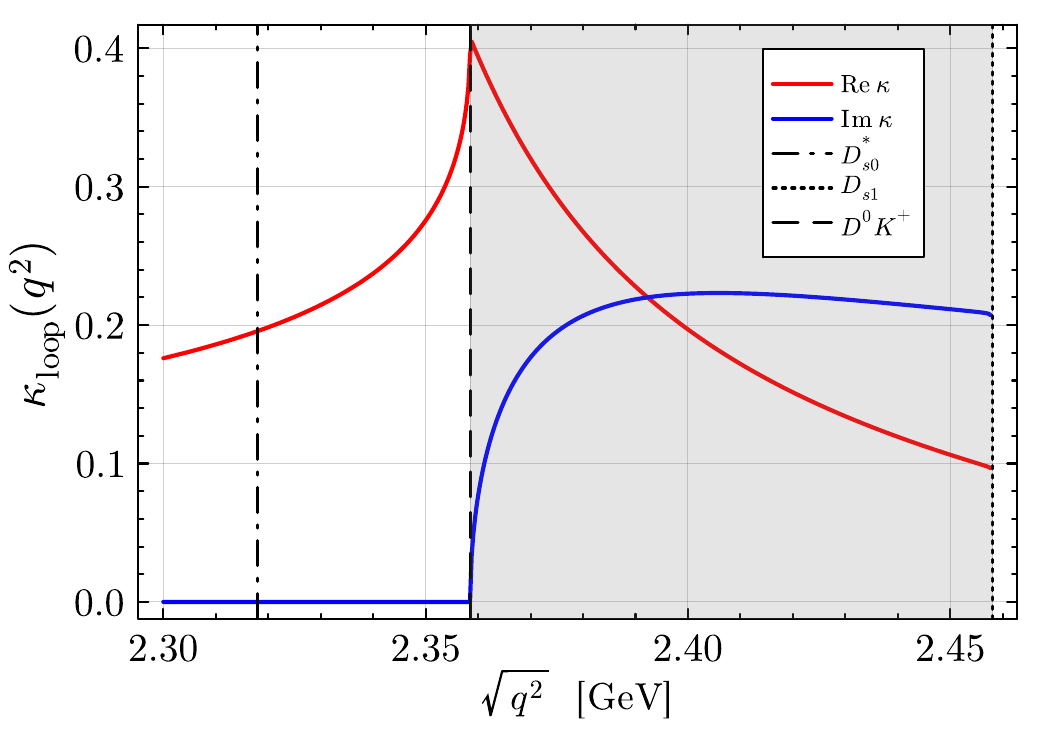}
\caption{Momentum dependence of the effective loop coupling $\kappa_{\rm{loop}}(q^2)$ in Eq.~\eqref{kappaloop}. For presentation purposes, here the loop integration in Eq.~\eqref{J0int} (see also Appendix~\ref{app:J}) is performed for the masses of $D^{(*)+}$ and $K^0$.
The vertical dash-dotted line shows the position
of $q^2=m_{\dszero}^2$ relevant for the two-body decay $\dsone(2460)\to\gamma\dszero(2317)$ (see Eqs.~\eqref{kappadef} and \eqref{kappaon}). The gray shaded region shows the
range of the phase space integration in $p_{12}^2=q^2$ in the three-body decay $\dsone(2460)\to\gamma D^0K^+$,
$(m_{D^{0}}+m_{K^{+}})^2\leqslant q^2\leqslant m_{\dsone}^2$ (see Eq.~\eqref{WidthDs1DKgamma}).
The plot for the decay $\dsone(2460)\to\gamma D^+K^0$ looks similar and is not shown. Note also that in the actual calculations performed in this work the spin-average masses in Eq.~\eqref{masses1} are used.}
\label{fig:kappaq2}
\end{figure}

Formally, the decay $\dsone(2460)\to \gamma\dszero(2317)$ can proceed through the diagrams shown in Fig.~\ref{fig:Ds1Ds0gamma},
so the amplitude $\kappa(q^2)$ in Eq.~\eqref{M1} acquires two contributions,
\be
\kappa(q^2)=\kappaoff+\kappa_{\rm cont},
\label{kappadef}
\ee
where we introduce an effective momentum-dependent contribution from the loop, $\kappaoff$, while the contact term $\kappa_{\rm cont}$ parametrises short-distance physics that is not captured by the molecular component. The latter can be estimated in a model-dependent way assuming a particular nature for such short-distance contributions---see, for example, Eq.~\eqref{kappacontesrimate} below for the estimate obtained in a $c\bar{s}$ model for the $\dsone(2460)$.
Note also that, within the effective field theory framework employed in this work, all couplings are real and complex phases may arise solely from intermediate particles going on shell---see, for example, Fig.~\ref{fig:kappaq2} were $\kappa_{\rm loop}(q^2)$ is demonstrated to develop a nonvanishing imaginary part in a certain kinematical regime.
In the molecular model for the positive-parity $D_{s0}^*(2317)$ and $D_{s1}(2460)$ mesons~\cite{Barnes:2003dj,Kolomeitsev:2003ac,Chen:2004dy,Guo:2006fu,Guo:2006rp,Gamermann:2006nm,Lutz:2007sk, Faessler:2007gv}, the loop contribution is sizeable, while
if both external hadrons were compact states, the
loops would be negligible. A key observation of this
work is that one can determine the significance of the
loop contribution by exploiting its $q^2$-dependence. Thus we start from evaluating $\kappaoff$. 

Assuming that both $\dsone(2460)$ and $D^{*}_{s0}(2317)$ are dynamically generated from the $D^*K$-$D_s^*\eta$ and $DK$-$D_s\eta$ coupled channels, respectively, within unitarised chiral perturbation theory (UChPT)~\cite{Kolomeitsev:2003ac,Guo:2006fu,Guo:2006rp} we employ the effective Lagrangians,
\begin{align}
{\cal L}^{(0)}_{\rm HM}=\frac{1}{\sqrt{2}}g_{DK}D_{s0}^{*}&\left(D^{+\dagger}K^{0\dagger}+D^{0\dagger}K^{+\dagger}\right)\nonumber\\
&+g_{D_{s}\eta}D^{*}_{s0}D^{+\dagger}_{s}\eta^{\dagger}+\mbox{h.c.},\label{LagHM0}\\
{\cal L}^{(1)}_{\rm HM}=\frac{1}{\sqrt{2}}g_{D^{*}K}D^{\mu}_{s1}&\left(D_{\mu}^{*+\dagger}K^{0\dagger}+D_{\mu}^{*0\dagger}K^{+\dagger}\right)\nonumber\\
&+g_{D^{*}_{s}\eta}D^{\mu}_{s1}D^{*+\dagger}_{s,\mu}\eta^{\dagger}+\mbox{h.c.},
\label{LagHM1}
\end{align}
with the coupling constants between hadronic molecular states and their constituent mesons determined from the residues of the $S$-wave scattering amplitudes in UChPT~\cite{Liu:2012zya,Fu:2021wde},
\be
\begin{split}
g_{DK}=(9.4\pm 0.3)~\mbox{GeV},\\
g_{D_{s}\eta}=(7.4\pm 0.1)~\mbox{GeV},\\
g_{D^*K}=(10.1^{+0.8}_{-0.9})~\mbox{GeV},\\
g_{D_s^*\eta}=(7.9\pm 0.3)~\mbox{GeV}.
\label{couplings}
\end{split}
\ee

Finally, the effective Lagrangian for the magnetic decays $V\to\gamma P$ reads~\cite{Amundson:1992yp,Hu:2005gf}\footnote{Here $v$ stands for the 4-velocity of the $D_{(s)}^*$ meson. However, in what follows we do not distinguish it from the 4-velocity of the $\dsone$ (see, for example, Eq.~\eqref{Lagcont} below) since the difference is of subleading order in the heavy quark mass expansion and can thus be disregarded.\label{ftnote}}
\be
\begin{split}
{\cal L}_{\rm MM}=\frac{i}{2}eF^{\mu\nu}\sqrt{m_{D}m_{D^{*}}} \left[\varepsilon^{\mu\nu\alpha\beta}v_{\alpha}\left(\beta Q+\frac{Q_c}{m_c}\right)_{ab}\right.\\
\left.\times(P_{a}V^{\dagger \beta}_{b}-V^{\beta}_{a}P^{\dagger}_{b})+\left(\beta Q-\frac{Q_c}{m_c}\right)_{ab} V^{\mu}_{a}V^{\dagger\nu}_{b}\right],
\label{LagMM}
\end{split}
\ee
where the heavy-quark spin multiplets are filled with the open-charm heavy--light pseudoscalar and vector mesons,
\be
P=(D^0,D^+,D_s^+),\quad V=(D^{*0},D^{*+},D^{*+}_s),
\ee
and the subscripts $a$ and $b$ label the light quark flavour. The parameter $\beta$ defines the contribution of the light quarks ($u$, $d$, and $s$) to the magnetic moment of the meson, with $Q=\mbox{diag}(2/3,-1/3,-1/3)$ for the matrix of the light-quark charges, while the respective contribution from the heavy charm quark
is provided by the term $Q_c/m_c$, with $Q_c=2/3$ and $m_c$ for the $c$-quark charge and mass, respectively.
The parameters $\beta$ and $m_c$ can be fixed directly from the experimental data on the radiative decays of the $D^*$ mesons; see, for example, Ref.~\cite{Hu:2005gf}.
It should be noted, however, that the numerical values of these parameters are very sensitive to the experimental inputs, which have changed appreciably in recent years. In addition, only the total width of the charged $D^*$ meson has been measured experimentally. Thus, in the current analysis, we fix the charm quark mass to a phenomenologically adequate value often adopted in the literature for its pole mass (see, for example, Refs.~\cite{Montana:2020var,Petreczky:2020tky,Beraudo:2022dpz,Huang:2024zni}) and extract the parameter $\beta$ from the data on the measured partial decay width $\Gamma(D^{*+}\to \gamma D^+)=(1.33\pm0.33)~\mbox{keV}$~\cite{ParticleDataGroup:2024cfk}. 
This yields
\be
m_c=1.5~\mbox{GeV},\quad
\beta^{-1}=360^{+27}_{-21}~\mbox{MeV},
\label{params}
\ee
with $\beta^{-1}$ of the order of $\Lambda_{\rm QCD}$, as expected.
For the masses of the $D$ meson and kaon, we use their isospin-averaged values,
\be
\begin{split}
m_D=\frac12(m_{D^0}+m_{D^+})=1867~\mbox{MeV},\\
m_K=\frac12(m_{K^0}+m_{K^+})=496~\mbox{MeV},
\label{masses1}
\end{split}
\ee
while for the other masses, we use their standard values quoted in the Review of Particle Physics~\cite{ParticleDataGroup:2024cfk}.

Notice that the $D_{(s)}^*$ propagator in the loop in Fig.~\ref{fig:Ds1Ds0gamma} is contracted with the photon emission vertex derived from the Lagrangian in Eq.~\eqref{LagMM}. Since this vertex contains a totally antisymmetric Levi-Civita tensor contracted with the $D_{(s)}^*$ 4-velocity (see footnote \ref{ftnote}), the longitudinal part of the $D_{(s)}^*$ polarisation tensor drops out. Thus, the loop integrals are ultraviolet convergent and no regularisation procedure needs to be invoked. In addition, the tiny widths of the mesons in the loop are disregarded for simplicity. Then, with the help of the effective Lagrangians in Eqs.~\eqref{LagHM0}, \eqref{LagHM1} and \eqref{LagMM}, an explicit expression for the effective coupling $\kappaoff$ 
introduced in Eq.~\eqref{kappadef} reads
\be
\begin{split}
&\kappaoff=\frac{e\sqrt{m_D m_{D^*}}}{12 m_c}
\Bigl[g_{DK}g_{D^*K}(\beta  m_c+4)\\
&\hspace*{0.1\columnwidth}\times J^{(0)}(m_{\dsone}^2,q^2,0,m_{D^*}^2,m_D^2,m_K^2)\\
&\hspace*{0.18\textwidth}-2g_{D_s\eta}g_{D^{*}_s\eta}(\beta  m_c-2)\\
&\hspace*{0.1\columnwidth}\times J^{(0)}(m_{\dsone}^2,q^2,0,m_{D^{*}_s}^2,m_{D_s}^2,m_{\eta}^2) \Bigr],
\label{kappaloop}
\end{split}
\ee
where $J^{(0)}$ is the standard scalar three-point loop function,
\begin{equation}
\begin{split}\label{J0int}
J^{(0)}(k_1^2,k_2^2,&k_3^2,m^2_{1},m^2_{2},m^2_{3}) \\
&=\frac{1}{16\pi^2}\int_0^1\delta(1-x_{1}-x_{2}-x_{3})\frac{d^3 x}{\Delta_{3}},
\end{split}
\end{equation}
with $x_{1,2,3}$ denoting the Feynman parameters and
\be
\Delta_{3}=\sum_{i=1}^3x_im_i^2-x_1x_2k_3^2-x_1x_3k_2^2-x_2x_3 k_1^2.
\ee
Further details of this calculation and the explicit expression for a generic loop integral $J^{(0)}$ are provided in Appendix~\ref{app:J}. 

\begin{figure*}[t]
    \centering
    \begin{tabular}{ccc}
    \includegraphics[width=0.3\textwidth]{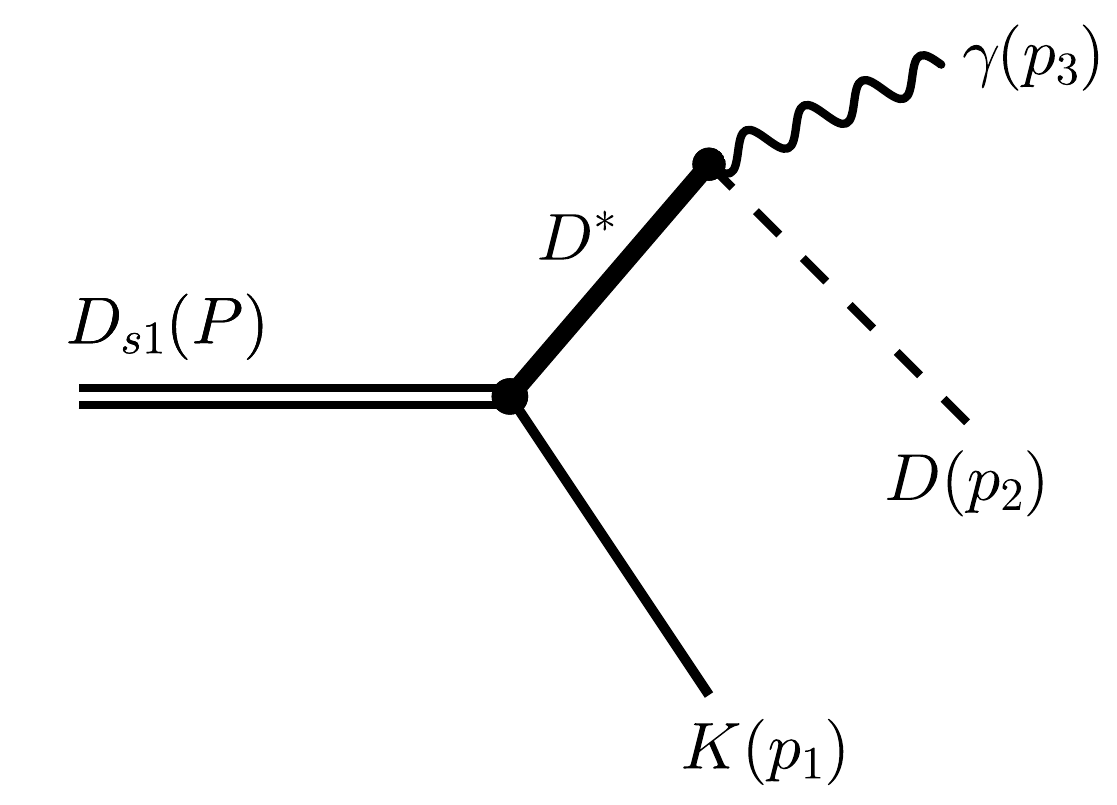}  & \includegraphics[width=0.3\textwidth]{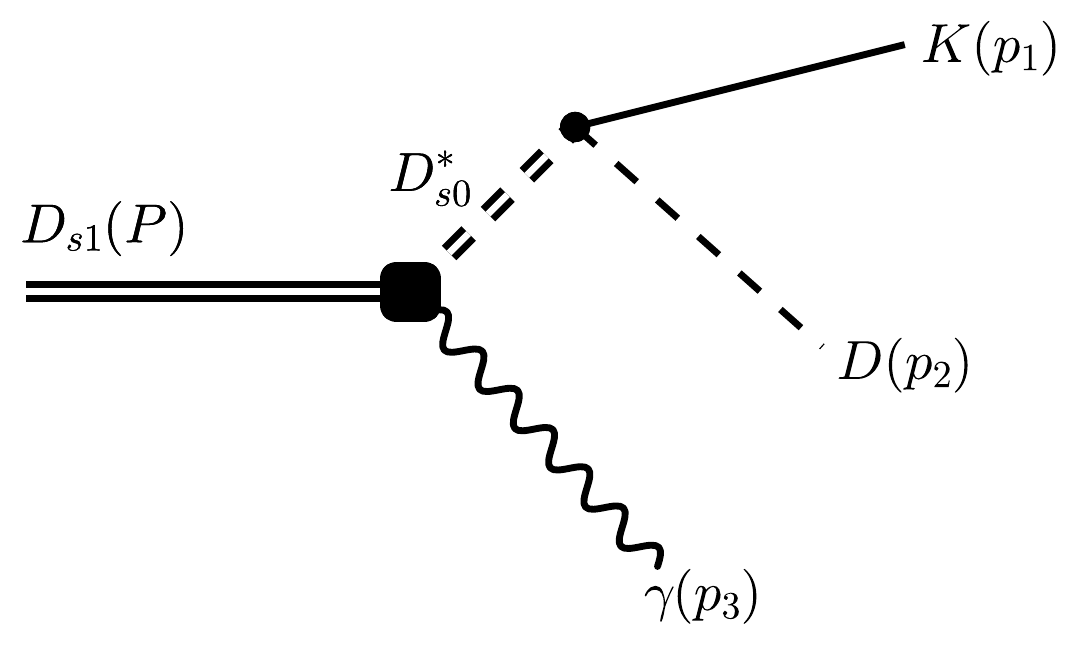} &
    \includegraphics[width=0.3\textwidth]{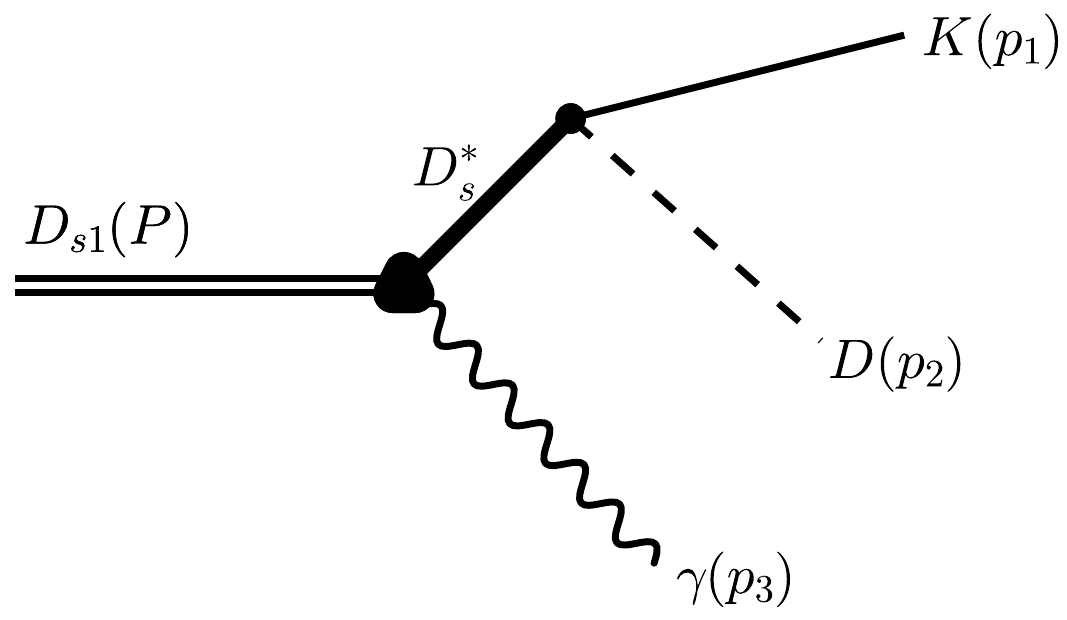}
    \\
     (a)  & (b) & (c)
    \end{tabular}
    \caption{Contributions to the decay amplitude $\dsone\to \gamma DK$ as given in Eq.~\eqref{chains}. The structure of the vertex $\dsone\to \gamma \dszero$ in diagram (b) is shown in Fig.~\ref{fig:Ds1Ds0gamma}. Details concerning the vertex $\dsone\to \gamma D_{s}^*$ in diagram (c) can be found in Refs.~\cite{Cleven:2013rkf,Cleven:2014oka,Fu:2021wde}.}
    \label{fig:Ds1DKgamma}
\end{figure*}

Then the real on-shell value $\kappaon$ entering Eq.~\eqref{kappadef} for the two-body radiative decay $\dsone\to\gamma\dszero$ is straightforwardly calculated employing the parameters listed in Eqs.~\eqref{couplings} and \eqref{masses1} to be
\be
\kappaon = 0.190\pm 0.004.
\label{kappaon}
\ee
The uncertainty here comes from that of $\beta$ quoted in Eq.~(\ref{params}).
Furthermore, in Fig.~\ref{fig:kappaq2}, we show the momentum dependence of the real and imaginary parts of $\kappaoff$ evaluated in a broad range of $q^2$ relevant for the further studies in this work. It is evident from this figure that the $q^2$-dependence in the near-threshold region, $q^2\sim (m_D+m_K)^2$, is rather pronounced. This behaviour arises from the two nearby singularities: (i)~the $DK$ threshold at $m_{D^0}+m_{K^+}=2.359$~GeV and (ii)~the triangle singularity~\cite{Landau:1959fi,Guo:2019twa} at $(2.319-i0.013)$~GeV, evaluated using the formula in Ref.~\cite{Bayar:2016ftu}.
This nontrivial $q^2$-dependence provides a near-threshold enhancement to the $DK$ invariant mass distribution for the three-body decays $D_{s1}(2460)\to \gamma DK$, which will be discussed in Sec.~\ref{sec:3body} below.

\begin{figure}[t!]
\centering
\includegraphics[width=0.95\columnwidth]{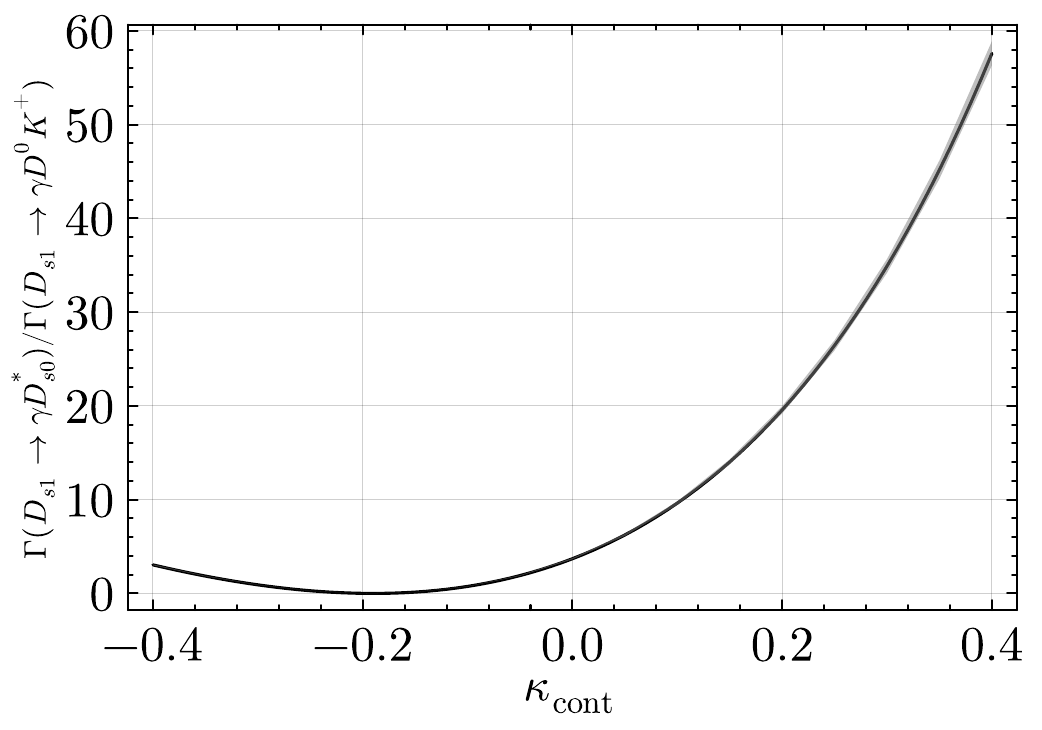}
\caption{The ratio ${\mathcal R}$ of the widths (branching fractions) for the radiative decays $\dsone(2460)\to \gamma D^{*}_{s0}(2317)$ and $\dsone(2460)\to \gamma D^0K^+$ as function of the contact interaction strength $\kappa_{\rm cont}$ in Eq.~\eqref{Lagcont}, varied over the natural range $[-0.4,0.4]$. The (nearly invisible by eye) band corresponds to the uncertainty in the determination of the contact parameter $\alpha_{\rm cont}$ as given in Eq.~\eqref{alphanum} with three times the standard deviation.}
\label{fig:ratio}
\end{figure}

The Lagrangian for the  leading-order contact interactions relevant for the radiative decays of the $D_{sJ}$ mesons reads~\cite{Lutz:2007sk,Cleven:2014oka,Fu:2021wde},
\be
\begin{split}
{\cal L}_{\rm cont}=&\, \alpha_{\rm cont} F_{\mu\nu}\Bigl(v^{\mu}D^{*}_{s0}D^{*\dagger\nu}_{s}+D^{\mu}_{s1}v^{\nu}D_{s}^{\dagger}\\
&\hspace*{0.4\columnwidth}+\varepsilon^{\mu\nu\alpha\beta}D_{s1\alpha}D^{*\dagger}_{s\beta}\Bigr)\\
&+\kappa_{\rm cont}\varepsilon^{\mu\nu\alpha\beta}F_{\mu\nu}v_{\beta}D_{s1\alpha}D^{*\dagger}_{s0}+\mbox{h.c.},
\label{Lagcont}
\end{split}
\ee
where $v$ denotes the 4-velocity of the $\dsone$ meson (see also footnote~\ref{ftnote} above), and $\kappa_\text{cont}$ has been introduced in Eq.~\eqref{kappadef}.
The first term in Eq.~\eqref{Lagcont},  which controls the radiative transitions from the $D_{sJ}$
to the ground state $D$-mesons, will be needed below for the calculation of the continuum transition; see
the last diagram in
Fig.~\ref{fig:Ds1DKgamma}.
The coupling $\alpha_{\rm cont}$ deserves a comment. Its value was previously
fixed from the averaged ratio of the branching fractions~\cite{ParticleDataGroup:2024cfk}
\be
R_2=\frac{\Br(\dsone(2460)\to \gamma D_s)}{\Br(\dsone(2460)\to \pi D_s^*)}
\label{R2}
\ee
and further employed to predict other ratios like 
\be
R_1=\frac{\Br(\dszero(2317) \to  \gamma D_s^*)}{\Br(\dszero(2317) \to \pi D_s)}
\label{R1}
\ee
in Refs.~\cite{Cleven:2014oka,Fu:2021wde}. Recently, Belle II announced the first observation of the radiative decay $\dszero(2317) \to \gamma D_s^*$ and a measurement of the ratio $R_1$
\cite{Belle-II:2025dzk}, which can be employed to update the extraction of $\alpha_{\rm cont}$ but demonstrates a tension with the theoretical prediction contained in Ref.~\cite{Fu:2021wde}. 
Furthermore, predictions for $R_1$ obtained in the molecular picture for the $\dszero(2317)$ and using $R_2$ from various experimental measurements~\cite{Belle:2003kup,Belle:2003guh,BaBar:2004yux} as input sizably differ from each other---see Appendix~\ref{app:R1} 
for a brief overview. We notice, however, that the dependence of the results of this work on the value of $\alpha_{\rm cont}$ is weak, since the contribution from the last diagram in Fig.~\ref{fig:Ds1DKgamma} is small numerically. Therefore, for the purposes of the present work, we perform a straightforward simultaneous fit to the experimental values of both aforementioned ratios of the branching fractions, $R_1=0.38\pm0.05$ (from the PDG FIT~\cite{ParticleDataGroup:2024cfk}) and $R_2=0.071\pm0.007$ (from Belle II~\cite{Belle-II:2025dzk}), to arrive at
\be
\alpha_{\rm cont}=-0.030\pm 0.008,
\label{alphanum}
\ee
which is used in the calculations below (see, in particular, Figs.~\ref{fig:Dalitzproj}-\ref{fig:ratio}).

\begin{figure}[t!]
\centering
\includegraphics[width=0.95\columnwidth]{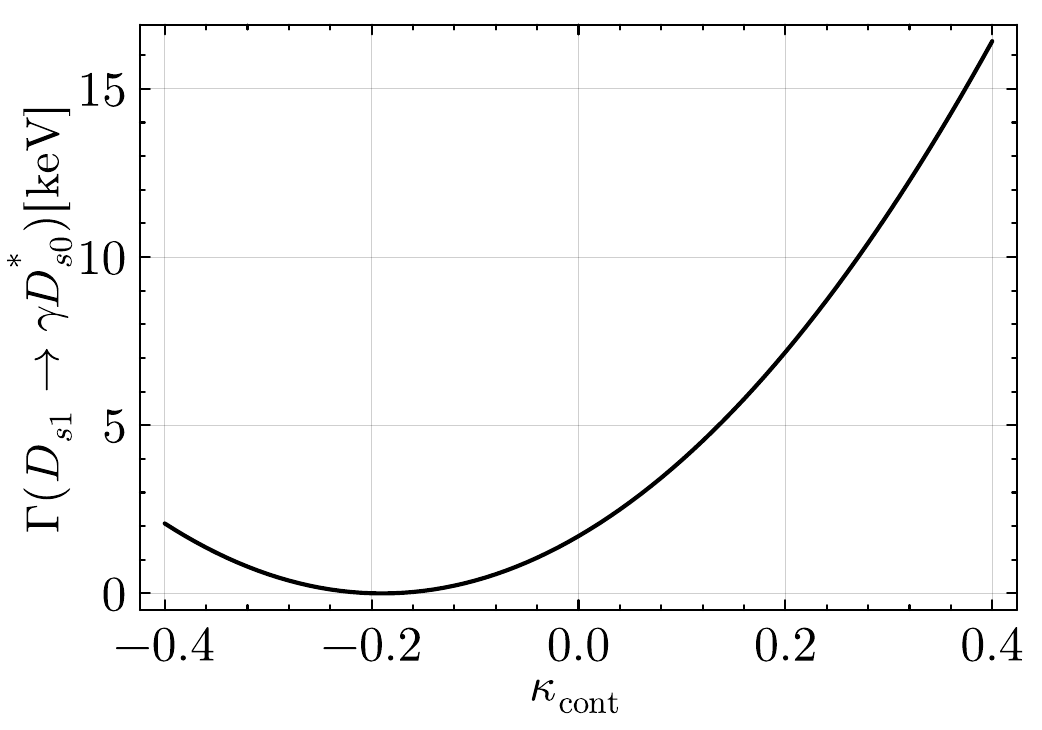}
\caption{Width of the radiative decay $\dsone(2460)\to\gamma \dszero(2317)$ in Eq.~\eqref{widthkappa} for the strength of the contact interaction $\kappa_{\rm cont}$ in Eq.~\eqref{Lagcont} varied in a natural range $[-0.4,0.4]$.}
\label{fig:kappa2}
\end{figure}

The magnetic coupling $\kappa_{\rm cont}$ that, in the studied decay $\dsone(2460)\to \gamma \dszero(2317)$, defines the strength of the contact diagram in Fig.~\ref{fig:Ds1Ds0gamma} is hitherto unknown, including its sign. This fact prevents us from making a definite prediction for the studied radiative decay, so the corresponding decay width in Eq.~\eqref{WidthDs1Gsogamma} can be presented in the form
\be
\begin{split}
\Gamma(\dsone(2460)&\to\gamma \dszero(2317))\\
&=47\times\bigl(0.190(4)+\kappa_{\rm cont}\bigr)^2~\mbox{keV}.
\label{widthkappa}
\end{split}
\ee
A simple order-of-magnitude estimate for $|\kappa_{\rm cont}|$ is given by the ratio $\Lambda_{\rm QCD}/m_c\simeq 0.2$, since the decay $\dsone(2460)\to \gamma \dszero(2317)$ involves a heavy quark spin flip. A model-dependent estimate for $\kappa_{\rm cont}$ can be obtained from the calculated radiative decay width of a charm-strange meson in a model that does not consider the molecular component.
In particular, applying the formula for the width in Eq.~\eqref{WidthDs1Gsogamma} to the result $\Gamma(1^+(c\bar{s})\to 0^+(c\bar{s})+\gamma)\approx 2.74$~keV obtained in the chiral doublet model~\cite{Bardeen:2003kt}, one readily finds
\be
\textcolor{blue}{|\kappa_{\rm cont}|\simeq 0.24,}
\label{kappacontesrimate}
\ee 
in good agreement with the order-of-magnitude estimate above.
Remarkably, comparing the above estimates with the result in Eq.~\eqref{kappaon}, we observe that, in line with the power counting provided in Ref.~\cite{Cleven:2014oka},
\be
|\kappa_{\rm cont}|\simeq|\kappaon|.
\ee
Therefore, the dependence of the width $\Gamma(\dsone(2460)\to\gamma \dszero(2317))$ in Eq.~\eqref{widthkappa} on the strength of the contact interaction $\kappa_{\rm cont}$, with $\kappaon$ fixed to the value in Eq.~\eqref{kappaon}, is rather pronounced---we plot it in Fig.~\ref{fig:kappa2} for $\kappa_{\rm cont}$ varied within a natural (and sufficiently broad) range $[-0.4,0.4]$ motivated by the estimates above.

The dependence in Fig.~\ref{fig:kappa2} allows for yet another estimate of the value of $\kappa_{\rm cont}$ based on the existing experimental data for the partial width of the radiative decay $\dsone(2460)\to\gamma \dszero(2317)$. The averaged experimental branching fraction for this decay is (though rather vaguely) known to be~\cite{ParticleDataGroup:2024cfk}
\be
\Br(\dsone\to\gamma \dszero)=3.7_{-2.4}^{+5.0}\%.
\label{Brexp}
\ee
However, only a very high upper limit has been established experimentally for the total $\dsone(2460)$ width \cite{ParticleDataGroup:2024cfk},
\be
\Gamma_{\rm tot}^{\rm exp}(\dsone)<3.5~\mbox{MeV}\quad (\mbox{CL}=95\%).
\label{widthexp}
\ee
To arrive at a more definite estimate for the partial decay width $\Gamma(\dsone\to\gamma \dszero)$, we sum up the known partial decay widths for the $\dsone(2460)$, as a hadronic molecule, into various final states collected in Table~\ref{tab:widths} to obtain
\be
\Gamma_{\rm tot}^{\rm th}(\dsone)\simeq 200~\mbox{keV}.
\label{widthth}
\ee
Therefore, for the central value of the branching fraction in Eq.~\eqref{Brexp}, we arrive at an estimate
\be
\Gamma(\dsone\to\gamma \dszero)\simeq 7~\mbox{keV},
\ee
and, employing the curve in Fig.~\ref{fig:kappa2}, find (including the sign)\footnote{We disregard the second solution that corresponds to a large and negative value of $\kappa_{\rm cont}$ lying beyond the phenomenologically adequate range shown in Fig.~\ref{fig:kappa2}.}
\be
\kappa_{\rm cont}\simeq 0.2,
\label{kappacont}
\ee
again in good agreement with the previous estimates. However, given the almost 100\% uncertainty in the averaged measured branching fraction in Eq.~\eqref{Brexp},
we are forced to conclude that the uncertainty of the result in Eq.~\eqref{kappacont} is at the level of the central value itself. Thus, although studies of the radiative decay $\dsone(2460)\to\gamma \dszero(2317)$ alone provide consistent estimates for the value of $\kappa_{\rm cont}$, such estimates are either model-dependent or come with a large uncertainty. The latter can be reduced by excluding the vaguely known partial decay width $\Gamma(\dsone\to\gamma \dszero)$ from consideration and resorting to ratios of branching fractions instead. In particular, below we calculate the widths of the three-body radiative decays $\dsone(2460)\to \gamma DK$ and argue that they can allow one to advance in extracting the short-range contribution $\kappa_{\rm cont}$ and probing the nature of the mesons $D^{*}_{s0}(2317)$ and $\dsone(2460)$.

\begin{table}[t!]
\caption{Partial decay widths of $\dsone(2460)$ into various final states obtained within the molecular model. We quote the central values of the results presented in the respective references. The asterisk indicates that the $D_s\pi^0\pi^0$ width was estimated from the $D_s\pi^+\pi^-$ one.}
\begin{ruledtabular}
\begin{tabular}{lcccccc}
Mode &$D_s^*\pi^0$ & $D_s\gamma$ &$D_s^*\gamma$ &$D_s\pi^+\pi^-$& $D_s\pi^0\pi^0$ & $\gamma D^{*}_{s0}$ \\
\hline
Width [keV] & 111 & 42 & 13 &16 & 8$^*$& $\simeq 1...10$ \\
Reference &\cite{Fu:2021wde}&\cite{Fu:2021wde}&\cite{Fu:2021wde}&\cite{Tang:2023yls}&\cite{Tang:2023yls}& Fig.~\ref{fig:kappa2}
\end{tabular}
\end{ruledtabular}
\label{tab:widths}
\end{table}

\section{The decays \texorpdfstring{$\dsone(2460)\to \gamma DK$}{Ds1toDKgamma}}
\label{sec:3body}

The amplitude of the three-body decay $\dsone(P)\to K(p_1) D(p_2)\gamma(p_3)$ takes a generic form
\be
\mathcal{M}(\dsone\to \gamma DK)=\mathcal{M}_{\mu\nu}(p_1,p_2,p_3)\epsilon^{\mu}(P)\epsilon^{*\nu}(p_{3}),
\label{Mampl}
\ee
where $\epsilon$ and $\epsilon^*$ are the polarisation vectors of the $\dsone$ and photon, respectively. The three main contributions to this amplitude come from the decay chains
\be
\begin{split}
&\mbox{(a)}\quad \dsone(2460)\to D^*K\to [\gamma D]K,\\
&\mbox{(b)}\quad \dsone(2460)\to \gamma \dszero\to \gamma [DK],\\
&\mbox{(c)}\quad \dsone(2460)\to \gamma D_s^*\to \gamma [DK],
\end{split}
\label{chains}
\ee
as depicted in Fig.~\ref{fig:Ds1DKgamma}.
In particular, the amplitude of the process $\dsone(2460)\to \gamma D^{*}_{s0}(2317)$ studied in the previous section enters as a building block in diagram (b) (see the effective vertex shown as a filled box) and brings in the dependence on the unknown parameter $\kappa_{\rm cont}$. For convenience, we also employ the shorthand notations
\be
p_{12}=p_1+p_2,\quad p_{23}=p_2+p_3
\label{p123}
\ee
for the momenta of the intermediate mesons in the diagrams in Fig.~\ref{fig:Ds1DKgamma}.
The various contributions to the tensor amplitude $\mathcal{M}_{\mu\nu}$ for the reactions $\dsone\to D^{*0}K^+\to \gamma D^0K^+$ and $\dsone\to D^{*+}K^0\to \gamma D^+K^0$, with all allowed contributions ($D^0K^+$, $D^+K^0$, and $D^+_{s}\eta$) included in the loop (see Fig.~\ref{fig:Ds1Ds0gamma}), take the form
\begin{align}
&\mathcal{M}_{\mu\nu}^{\rm a}(\gamma D^0K^+) \notag\\
&=-e\frac{\sqrt{2m_D m_{D^*}}}{3 m_c}\varepsilon_{\mu \nu\alpha\beta}\;p_3^{\alpha} v^{\beta} g_{D^* K} \left(\beta  m_c+1\right)G_{D^{*0}}(p_{23}), \notag\\[2mm]
&\mathcal{M}_{\mu\nu}^{\rm a}(\gamma D^+K^0) \notag \\
&=-e\frac{\sqrt{2m_D m_{D^*}}}{6 m_c}\varepsilon_{\mu\nu\alpha\beta}\;p_3^{\alpha} v^{\beta} g_{D^* K} \left(\beta  m_c-2\right) G_{D^{*+}}(p_{23}), \notag \\[2mm]
&\mathcal{M}_{\mu\nu}^{\rm b}(\gamma DK)\notag\\
&=-\sqrt{2}g_{DK}\left[\kappa_{\rm loop}(p^2_{12})+\kappa_{\rm cont}\right]\varepsilon_{\mu \nu\alpha\beta}\; p_3^{\alpha} v^{\beta} G_{D^{*}_{s0}}(p_{12}).
\label{ampls}
\end{align}
In the expressions above, the superscripts ``a'' and ``b'' indicate the contributions from diagrams (a) and (b) in Fig.~\ref{fig:Ds1DKgamma}, respectively.
A detailed description of the amplitude $\mathcal{M}_{\mu\nu}^{\rm c}$ for diagram (c) can be found in Ref.~\cite{Cleven:2013rkf}, so we refrain from quoting it here. Notice that, in all expressions above, the $D^*$ propagator (also the $D_s^*$ propagator in the skipped amplitude $\mathcal{M}_{\mu\nu}^{\rm c}$) is multiplied by the photon emission vertex derived from the Lagrangian in Eq.~\eqref{LagMM}. Since the photon vertex contains a totally antisymmetric Levi-Civita tensor contracted with the $D_{(s)}^*$ 4-velocity, the longitudinal part of the $D_{(s)}^*$ polarisation tensor drops out
(it is, however, retained in the sum over the $D^*$ polarisations in $\overline{|{\cal M}|^2}$ in Eq.~\eqref{WidthDs1DKgamma} below). Thus, for the propagators of the intermediate vector mesons in Fig.~\ref{fig:Ds1DKgamma}, we resort to a universal Breit--Wigner distribution,
\be
G_V^{-1}(p_{23})=p_{23}^2-m_V^2+im_V\Gamma_V,
\label{GX}
\ee
with $V=D^{*0}$, $D^{*+}$, and $D_s^*$. We use the standard values of the $D^*_{(s)}$ masses \cite{ParticleDataGroup:2024cfk} and for their widths we employ\footnote{Since the $D^{*0}$ width has not been measured yet, its value is taken from Ref.~\cite{Guo:2019qcn}, where it is evaluated from the $D^{*+}$ width using isospin symmetry.}
\be
\Gamma_{D^{*0}}=55.3~\mbox{keV},\quad\Gamma_{D^{*+}}=83.4~\mbox{keV},
\label{widths}
\ee
while the tiny width of $D^*_s$ \cite{Wang:2019mhm,Wang:2025fzj} is set to zero.

\begin{figure*}[t]
\centering
\includegraphics[width=0.47\linewidth]{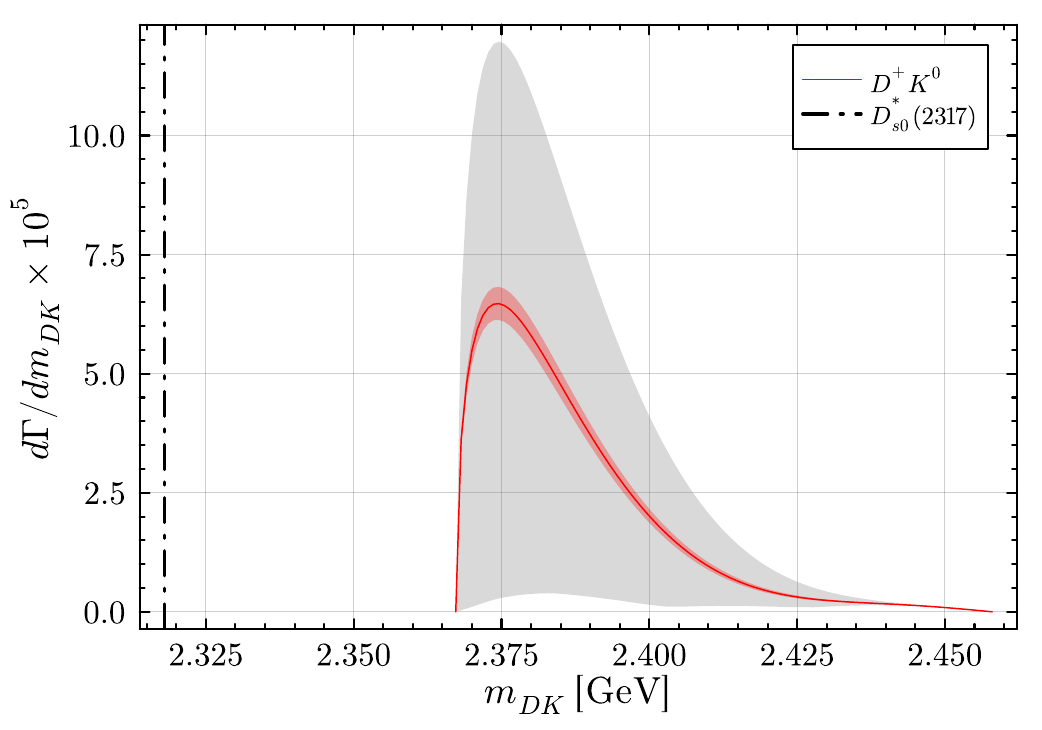}
\includegraphics[width=0.47\linewidth]{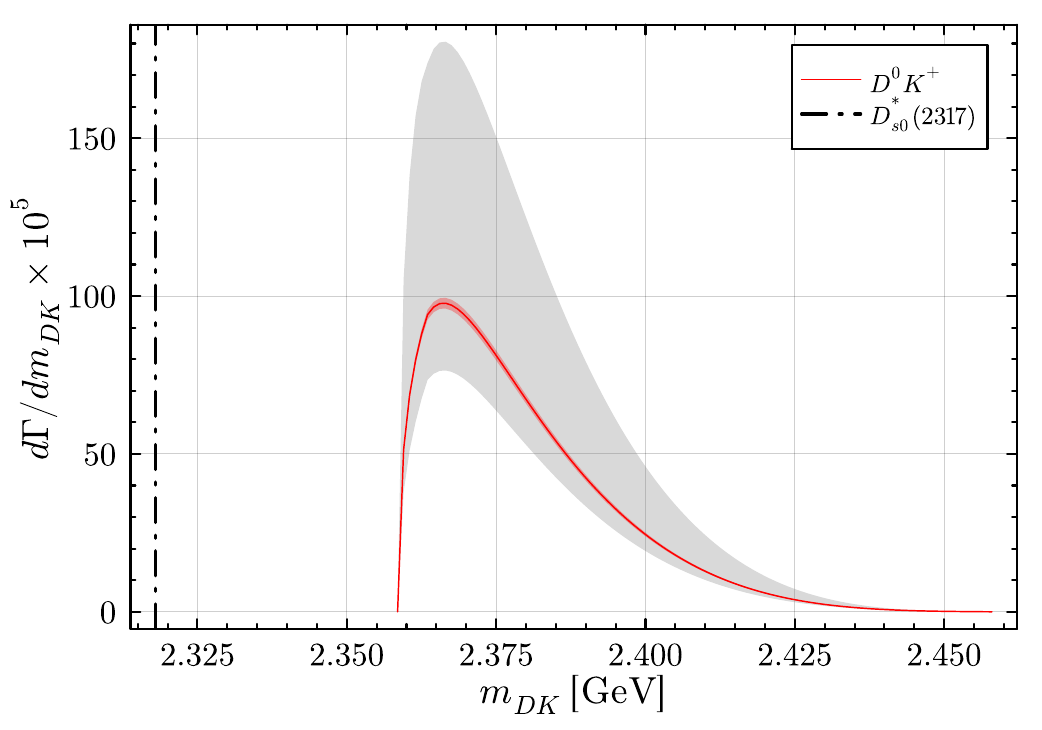}
\caption{$DK$ invariant mass distributions for the three-body radiative decay $\dsone(2460)\to \gamma D^+K^0$ (left) and $\dsone(2460)\to \gamma D^0K^+$ (right) obtained from Eq.~\eqref{WidthDs1DKgamma} upon partial integration over the phase space of the final state. In both plots, the red curve corresponds to $\kappa_{\rm cont}=0.2$ as suggested by Eq.~\eqref{kappacont} and the red band around it comes from the uncertainty in the determination of the contact parameter $\alpha_{\rm cont}$ as given in Eq.~\eqref{alphanum}; in both cases we use three times the corresponding standard deviation for the uncertainty of $\alpha_{\rm cont}$ to increase its visibility.  The gray bands correspond to $\alpha_{\rm cont}$ fixed to its central value in Eq.~\eqref{alphanum} and $\kappa_{\rm cont}$ varied in the range $[-0.4,0.4]$. }
\label{fig:Dalitzproj}
\end{figure*}

\begin{figure}[t!]
\centering
\includegraphics[width=0.95\columnwidth]{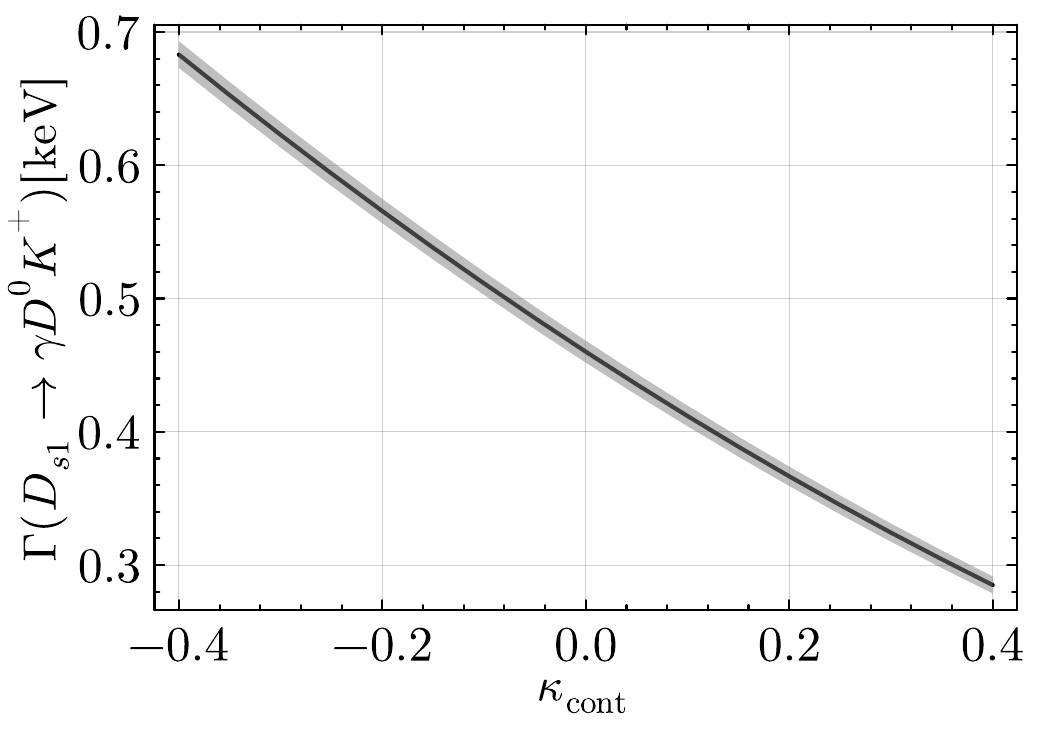}
\caption{Partial width of the radiative decay $\dsone(2460)\to \gamma D^0K^+$ as a function of the contact interaction strength $\kappa_{\rm cont}$ in Eq.~\eqref{Lagcont}, varied over the natural range $[-0.4,0.4]$. The narrow band corresponds to the uncertainty in the determination of the contact parameter $\alpha_{\rm cont}$ as given in Eq.~\eqref{alphanum} with three times the standard deviation. 
}
\label{fig:kappa3}
\end{figure}

For the $D^{*}_{s0}$ propagator entering the amplitude for the diagram in Fig.~\ref{fig:Ds1DKgamma}(b), we employ a Flatt{\'e} distribution~\cite{Flatte:1976xu},
\be
G_{D^{*}_{s0}}^{-1}(p_{12})=p_{12}^2-m_{D^{*}_{s0}}^2+\frac{g_{DK}^2}{8\pi\sqrt{p_{12}^2}}(\gamma+ik)+im_{D^{*}_{s0}}\Gamma_{D^{*}_{s0}},
\label{eq:FLG}
\ee
where $\gamma=-ik(m^2_{D^{*}_{s0}})$
and the momentum $k$ in the $D^{*}_{s0}$ centre-of-mass frame reads
\be
k(p_{12}^2)=\frac{1}{2\sqrt{p_{12}^2}}\lambda^{1/2}(p_{12}^2,m_D^2,m_K^2),
\ee
with
\be
\lambda(a,b,c)=a^2+b^2+c^2-2ab-2ac-2bc
\label{triangle}
\ee
being the standard K{\"a}ll\'en triangle function. For the real part of the pole, interpreted as the $D^{*}_{s0}$ nominal mass, we use
\be
m_{D^{*}_{s0}}=2317~\mbox{MeV},
\ee
and for the $D^{*}_{s0}$ width we take the value \cite{Liu:2012zya}
\be
\Gamma_{D^{*}_{s0}}=132~\mbox{keV}.
\ee
Finally, the differential decay width,
\be
d\Gamma(\dsone\to \gamma DK)=\frac{1}{(2\pi)^3}\frac{\overline{|\mathcal{M}|^2}}{32m_{\dsone}^3}dp_{12}^2dp_{23}^2,
\label{WidthDs1DKgamma}
\ee
is integrated over the phase space of the three-body final state. 
In particular, the differential widths 
$d\Gamma(\dsone(2460)\to\gamma D^+K^0)/dm_{DK}$ and $\Gamma(\dsone(2460)\to\gamma D^0K^+)/dm_{DK}$, with $m_{DK}=\sqrt{p_{12}^2}$,
are shown in Fig.~\ref{fig:Dalitzproj}. An order-of-magnitude difference between the two widths should not come as a surprise given a strong cancellation between the contributions to the $D^{*+}$ magnetic moment from the charm quark and the cloud of light quarks, so this suppression has the same origin as the relation $\Gamma(D^{*+}\to\gamma D^+)\ll \Gamma(D^{*0}\to\gamma D^0)$. Thus, in what follows, we focus on the final state $\gamma D^0K^+$. In particular, the dependence of the total decay width $\Gamma(\dsone(2460)\to\gamma D^0K^+)$ on the short-range contact parameter $\kappa_{\rm cont}$ is shown in Fig.~\ref{fig:kappa3}.

\section{Discussion}
\label{sec:discussion}

The estimate for $\kappa_{\rm cont}$ in Eq.~\eqref{kappacont} suggests that positive values are more natural. 
We further notice that, for positive values of $\kappa_{\rm cont}$, the 
dependencies of the partial radiative decay widths $\Gamma(\dsone(2460)\to\gamma \dszero(2317))$ and $\Gamma(\dsone(2460)\to\gamma D^0K^+)$ shown
in Figs.~\ref{fig:kappa2} and \ref{fig:kappa3}, respectively, exhibit pronounced but opposite patterns: while the curve in Fig.~\ref{fig:kappa2} rises, the one in Fig.~\ref{fig:kappa3} falls.
Therefore, studying the ratio of the branching fractions defined in Eq.~\eqref{ratioR} appears to be advantageous from both the theoretical and experimental points of view. Indeed, the predicted dependence ${\cal R}(\kappa_{\rm cont})$ depicted in Fig.~\ref{fig:ratio} demonstrates a rapid rise with increasing $\kappa_{\rm cont}$---the values of ${\cal R}$ predicted for $\kappa_{\rm cont}\simeq 0$ and $\kappa_{\rm cont}\simeq 0.4$ differ by an order of magnitude. 
From the experimental point of view, measuring a ratio of two branching fractions should be much easier than determining the absolute values of the corresponding partial decay widths separately. A measurement of either the partial widths of both radiative decays
$\dsone(2460)\to\gamma \dszero(2317)$ and $\dsone(2460)\to\gamma D^0K^+$, or at least the ratio of their branching fractions ${\cal R}$ in Eq.~\eqref{ratioR},  with sufficient precision, will allow for a reliable determination of the short-range parameter $\kappa_{\rm cont}$. Moreover, given the specific pattern exhibited by the curve in Fig.~\ref{fig:ratio}, an experimentally established sufficiently high lower bound on the ratio ${\cal R}$ may already allow one to impose a restrictive constraint on the value of $\kappa_{\rm cont}$. 

A comment on the nature of the $D^{*}_{s0}(2317)$ meson and its influence on the studied three-body radiative decays is also in order here. Throughout this paper, we treated $D^{*}_{s0}(2317)$ as mainly a $DK$ molecular state, so its coupling to the corresponding channel is large---see the value of $g_{DK}$ in Eq.~\eqref{couplings}. Consequently, the amplitude of Fig.~\ref{fig:Ds1DKgamma}(b), which is directly proportional to this coupling, brings about a noticeable dependence of the total three-body decay width $\Gamma(\dsone\to \gamma D^0K^+)$ on the unknown parameter $\kappa_{\rm cont}$ that is in the spotlight of this investigation.
On the contrary, in the opposite limit of a purely compact $D^{*}_{s0}(2317)$, when its coupling to $DK$ is very much reduced, the contribution in Fig.~\ref{fig:Ds1DKgamma}(b) is negligible. Then the
amplitudes (a) and (c) in Fig.~\ref{fig:Ds1DKgamma} are parameter free and provide a prediction for the total width of the corresponding three-body decay,
\be
[\Gamma(\dsone(2460)\to \gamma D^0K^+)]_{{\rm compact }D^{*}_{s0}}\simeq 0.68~\mbox{keV}.
\ee
As can be concluded from Fig.~\ref{fig:kappa3}, to achieve this value in the molecular model for the $D^{*}_{s0}(2317)$ one would need $\kappa_{\rm cont}\simeq -0.4$ , which is negative and, in addition, appears unnaturally large in absolute value. This result reflects the fact
that the loop contribution encoded in $\kappaoff$ is sizable; see Fig.~\ref{fig:kappaq2}.

 We conclude, therefore, that the radiative decays of the $\dsone(2460)$  constitute a paradigmatic case of the reactions which, on the one hand, are sensitive to the short-range component of the decaying meson wave function and, on the other hand, allow one to quantify the latter short-range contribution in a model-independent way employing the experimental data. We further argue that the
experimental studies of the three-body radiative decay $\dsone(2460)\to\gamma D^0K^+$ may provide valuable insights into the nature of the enigmatic mesons $\dsone(2460)$ and $\dszero(2317)$. 
It follows from the results of this work that the branching fraction for this three-body radiative decay can be estimated at the level $10^{-3}$ (optimistically, at a per cent level)---see Fig.~\ref{fig:kappa3} for the most natural values of this partial decay width gained for $\kappa_{\rm cont}\simeq 0.1..0.2$ and the estimate for the $\dsone(2460)$ total width in Eq.~\eqref{widthth}. Given the present active phase of data taking at Belle II and a fast growth of the collected data sample, the experimental studies of the two- and three-body radiative decays of the $\dsone(2460)$ may become feasible already in the near future.

\begin{acknowledgements}
This work is supported in part by the National Key R\&D Program of China under Grant No. 2023YFA1606703; by the National Natural Science Foundation of China (NSFC) under Grants No. 12125507, No. 12361141819, and No. 12447101; and by the Chinese Academy of Sciences (CAS) under Grant No.~YSBR-101. A.N. would like to thank Alex Bondar for bringing his attention to the problem of radiative decays of the $\dsone$ mesons and Roman Mizuk for discussions of the Belle II capabilities in studies of such decays.
Work of A.N. was supported by Deutsche Forschungsgemeinschaft (Project No. 525056915). A.N. and C.H. also acknowledge the support from the CAS President’s International Fellowship Initiative (PIFI) (Grants No.~2024PVA0004\_Y1 and No.~2025PD0087, respectively).
\end{acknowledgements}

\appendix

\section{Loop integrals}
\label{app:J}

In this Appendix, we provide the analytical expressions
that appear in the amplitudes of the decays $\dszero\to\gamma D^*_{s}/D^*_{s0}$; see
the loop diagram in Fig.~\ref{fig:Ds1Ds0gamma} and
diagrams (b) and (c) in Fig.~\ref{fig:Ds1DKgamma}.
In particular, after tensor reduction of the corresponding three-point one-loop amplitudes \cite{Cleven:2013rkf}, we employ Package-X \cite{Patel:2015tea} to express the remaining scalar integrals in terms of PolyLog functions. The result reads
\be
\begin{split}
&J^{(0)}(M^2,q^2,0,m_1^2,m_2^2,m_3^2)\\
&=-\frac{1}{(4\pi)^2}\frac{1}{M^2-q^2}\bigg[\\
&\hspace*{0.045\columnwidth}\textrm{Li}_{2}\left(F_1^{(+)}+\textrm{sgn}(f_{1})i\varepsilon\right)
+\textrm{Li}_{2}\left(F_1^{(-)}-\textrm{sgn}(f_{1})i\varepsilon\right)\\
&-\textrm{Li}_{2}\left(F_2^{(+)}-\textrm{sgn}(f_{2})i\varepsilon\right)
-\textrm{Li}_{2}\left(F_2^{(-)}+\textrm{sgn}(f_{2})i\varepsilon\right)\\
&-\textrm{Li}_{2}\left(F_3^{(+)}-\textrm{sgn}(f_{1})i\varepsilon\right)
-\textrm{Li}_{2}\left(F_3^{(-)}+\textrm{sgn}(f_{1})i\varepsilon\right)\\
&+\textrm{Li}_{2}\left(F_4^{(+)}
-\textrm{sgn}(f_{3})i\varepsilon\right)
+\textrm{Li}_{2}\left(F_4^{(-)}+\textrm{sgn}(f_{3})i\varepsilon\right)\\
&-\textrm{Li}_{2}\left(F_5\right)
+\textrm{Li}_{2}\left(F_6\right)\bigg],
\end{split}
\label{Jint}
\ee
where
\begin{align}
&F_1^{(\pm)}=\frac{2q^2\delta_{12}-2M^2\delta_{13}}{M^2(M^2-\Delta_{1})+\delta_{12}q^2\pm (M^2-q^2)\lambda_1^{1/2}}, \notag\\
&F_2^{({\pm})}=\frac{2M^2(M^2-\Delta_2)+2\delta_{12}q^2}{M^2(M^2-\Delta_{1})\delta_{12}q^2\pm (M^2-q^2)\lambda_1^{1/2}},\notag\\
&F_3^{(\pm)}=\frac{2M^2\delta_{13}-2\delta_{12}q^2}{M^2\Delta_{2}-q^2\Delta_{1}\pm (M^2-q^2)\lambda_2^{1/2}},\notag\\
&F_4^{(\pm)}=\frac{2M^2\Delta_{2}-2(q^2-\delta_{12})q^2}{M^2\Delta_{2}-q^2\Delta_{1}\pm(M^2-q^2)\lambda_2^{1/2}},\\
&F_5=\frac{\delta_{23}\left(M^2\Delta_{2}+q^2(q^2-\delta_{12})\right)}{m_3^2M^4+BM^2+Dq^2},\notag\\
&F_6=\frac{\delta_{23}\left(M^2(\Delta_{2}-M^2)-q^2\delta_{12}\right)}{m_3^2M^4+BM^2+Dq^2} ,\notag
\end{align}
with
\be
\lambda_{1}=\lambda(M^2,m_1^2,m_2^2),\quad\lambda_{2}=\lambda(m_1^2,m_3^2,q^2),
\ee
and the triangle function $\lambda(a,b,c)$ defined in Eq.~\eqref{triangle}, while the signs of the functions $f_n$ ($n=1..3$),
\be
\begin{split}
f_{1}&=(M^2-q^2)(M^2\delta_{13}-q^2\delta_{12}),\\
f_{2}&=(M^2-q^2)(M^4-M^2\Delta_2+\delta_{12}q^2),\\\
f_{3}&=(M^2-q^2)(M^2\Delta_{1}+q^4-q^2\delta_{12}),
\end{split}
\ee
determine the Riemann sheet of $\textrm{Li}_2(z\pm i\varepsilon)$ for $\text{Re}(z)>1$.
Finally, the auxiliary functions are defined as
\be
\begin{split}
&\Delta_1=m_1^2+m_2^2-2m_3^2+q^2,\\
&\Delta_{2}=m^2_{1}-m_3^2+q^2,\\
&\delta_{ij}=m^2_{i}-m_j^2,\\
&B=m_3^4+m_1^2\delta_{23}-m_3^2q^2-m_2^2q^2-m_2^2m_3^2,\\
&D=-m_1^2\delta_{23}+m_2^2(q^2+\delta_{23}).
\end{split}
\ee

Note that, in the two-body decay studied in Sec.~\ref{sec:2body},
the loop integral in Eq.~\eqref{Jint} is evaluated at a fixed value $q^2=m^2_{\dszero}$, while in the three-body decays in Sec.~\ref{sec:3body}, it is evaluated at $q^2=(P-p_3)^2$, which varies across the three-body phase space.

\begin{table}[tbh]
\centering
\caption{Extracted values of $\alpha_{\rm cont}$ and the corresponding ratios $R_1$ predicted in the molecular picture obtained by taking individual measurements for $R_2$ from Belle~\cite{Belle:2003kup, Belle:2003guh} and BaBar~\cite{BaBar:2004yux} as input.}
\begin{ruledtabular}
\begin{tabular}{l|cc}
    Input $R_2$ & $\alpha_{\rm cont}$ & Predicted $R_1$ \\\hline
$0.55\pm0.15$~\cite{Belle:2003kup} &$\phantom{+}0.05\pm0.03$  & $0.010\pm0.001$ \\
$0.38\pm0.12$~\cite{Belle:2003guh} &  $\phantom{+}0.01\pm0.03$& $0.026\pm0.003$ \\
$0.274\pm0.049$~\cite{BaBar:2004yux} &$-0.01\pm0.02$  & $0.044\pm0.004$ \\
\end{tabular}
\end{ruledtabular}
\label{tab:R1}
\end{table}

\section{$R_1$ in the molecular picture}
\label{app:R1}

In this appendix, we discuss the ratio $R_1$ defined in Eq.~\eqref{R1}
employing the molecular picture. The numerical value of $R_1$ in Ref.~\cite{Fu:2021wde} was obtained using $\alpha_{\rm cont}$ fixed from the PDG FIT value $0.38\pm0.05$~\cite{ParticleDataGroup:2024cfk} for the ratio $R_2$ in Eq.~\eqref{R2}.
However, we note that the values of $R_2$ from different measurements differ sizably.
In Table~\ref{tab:R1}, we list the values of $\alpha_{\rm cont}$ and the corresponding predicted ratios $R_1$ obtained in the molecular picture by taking individual measurements for $R_2$ from Belle~\cite{Belle:2003kup,Belle:2003guh} and BaBar~\cite{BaBar:2004yux} as input. 
Clearly, the values of $R_1$ found in this way from the different 
experiments are not consistent and deviate from each other even more than the input quantities derived from $R_2$, since the contact term
interferes with the loop contribution.

\bibliography{refs}

\end{document}